\documentclass[12pt]{iopart}

\usepackage{iopams}
\usepackage{amssymb}
\usepackage{enumerate}
\usepackage{amsthm}
\usepackage{graphicx}
\usepackage{bm}
\usepackage[utf8]{inputenc}
\usepackage{float}

\begin{document}

\title[Deterministic physical systems under uncertain initial conditions]{Deterministic physical systems under uncertain initial conditions: the case of maximum 
entropy applied to projectile motion}
\author{Alejandra Montecinos$^1$, Sergio Davis${^{1,}}{^2}$, Joaqu\'in Peralta$^1$}
\address{$^1$Universidad Andres Bello, Departamento de Ciencias F\'{\i}sicas, Sazi\'e 2212 piso 7,
	Santiago, Chile}
\address{$^2$Comisi\'on Chilena de Energ\'{\i}a Nuclear, Casilla 188-D, Santiago, Chile}
\ead{alejandramontecinos@unab.com}

\begin{abstract}
The kinematics and dynamics of deterministic physical systems have been a foundation of our understanding of the world
since Galileo and Newton. For real systems, however, uncertainty is largely present via external forces such as friction or lack of
precise knowledge about the initial conditions of the system. In this work we focus in the latter case and describe the
use of inference methodologies in solving the statistical properties of classical systems subject to uncertain initial
conditions. In particular we describe the application of the formalism of Maximum Entropy (MaxEnt) inference to the problem of 
projectile motion given information about the average horizontal range over many realizations. By using MaxEnt we can
invert the problem and use the provided information on the average range to reduce the original uncertainty in the
initial conditions, while also achieving additional insights based on the shape of the posterior probabilities for the initial 
conditions probabilities and the projectile path distribution itself. The wide applicability of this procedure, as well
as its ease of use, reveals a useful tool by which to revisit a large number of physics problems, from classrooms to frontier research.
\end{abstract}
\date{\today}
\submitto{\EJP}
\maketitle

\section{Introduction}

Classical mechanical systems are completely deterministic, following well-established equations of motion such as
Newton's second law or Hamilton equations. However, uncertainty still is introduced in the lack of control of the
initial conditions of a mechanical system. Usually mechanical systems are chaotic, that is, highly sensitive to these
initial conditions, with the corresponding loss of predictability. Given a certain statistical distribution of initial
conditions, the result is of course a (non-trivial) statistical distribution of outcomes, obtained by propagating the
system forwards in time according to the equations of motion. This suggests the opposite situation: given statistical
properties of the outcomes of a classical system, can we obtain some information about the initial conditions that
produced these outcomes? This is an inverse problem from the point of view of probability theory, which can in principle
be addressed using Bayes' theorem~\cite{Jaynes2003} and/or information theoretical methods~\cite{CoverThomas2006}.

Since the time of Galileo Galilei, projectile motion has been a widely studied phenomenon from a large set of different
approaches. These range from including the presence of a resisting medium~\cite{Stewart2012}, to highly accurate analytical functions~\cite{Turkylmazoglu2016}. 
This particular motion presents an almost endless list of uses in different technological areas, such as ballistics~\cite{Hayen2003, Stewart2005, Morales2005, 
Parker1977}, sports~\cite{Erlichson1983, Linthorne2006}, rockets among others~\cite{Williams2013, Borghi2013}. In spite of the large 
body of information on projectile motion we are not aware of previous work where it is treated as an inverse problem in which 
information at later times is known and we infer the initial conditions. 

In this work we propose and solve the problem of a large number of realizations of projectile motion with known average horizontal 
range $\bar{R}$, where we infer the most unbiased probability distribution of initial angle $\theta_0$ and speed $v_0$. This
is achieved by the use of the Maximum Entropy principle (MaxEnt).

Our application to projectile motion is only a guide to the proposed procedure, and its choice is based fundamentally in 
strengthening statistical methods in students. By understanding the theoretical foundations that allow the solution of
this kind of problems, students will be able to diversify their applicability to many other problems known in physics.

\section{The principle of Maximum Entropy}

The principle of maximum entropy~\cite{Jaynes1957} (MaxEnt for short), proposed by Edwin T. Jaynes in 1957, is
extensively used as a general tool of probabilistic inference~\cite{Presse2013} applicable to numerical data analysis,  
image reconstruction, and inverse problems in general. In its modern formulation, MaxEnt establishes than the most 
unbiased probability distribution $P$ given some state of knowledge $I$ is the one that maximizes the Shannon-Jaynes 
entropy $S[P]$ (known also as relative entropy or the negative of the Kullback-Leibler divergence), given
by~\cite{Sivia2006}

\begin{equation}\centering
S(I_0 \rightarrow I) = - \int d\bm{x} P(\bm{x}|I)\ln \frac{P(\bm{x}|I)}{P(\bm{x}|I_0)},
\label{shannon}
\end{equation}
while being consistent with said knowledge. Here $\bm{x}=(x_1, \ldots, x_N)$ is the $N$-component
vector of continuous degrees of freedom of the system and $P(\bm{x}|I_0)$ represents a prior probability distribution, used as 
a starting point for the inference.

By maximizing $S(I_0 \rightarrow I)$ in Eq. \ref{shannon}, we obtain the probabilistic model that contains the least amount of 
information (i.e. the least biased) while reproducing the features one demands of it. Consider a constraint on the known expectation 
$f_0$ of a function $f(\bm{x})$ (this knowledge is represented by $I$). According to MaxEnt, the most unbiased model is given by 
maximizing $S$ in Eq. \ref{shannon} with the constraint $\left< f(\bm{x})\right> = f_0$. The constrained maximization
problem is usually implemented by means of a Lagrange multiplier $\lambda$, which leads to the well-known maximum entropy model,

\begin{equation}\centering
P(\bm{x}|\lambda) = \frac{1}{Z(\lambda)} \exp{(-\lambda f(\bm{x}))}P(\bm{x}|I_0),
\label{maxent}
\end{equation}

\noindent
with $$Z(\lambda)=\int d\bm{x}\exp{(-\lambda f(\bm{x}))}P(\bm{x}|I_0)$$ 
the partition function, and the Lagrange multiplier $\lambda$ obtained by the constraint equation 

\begin{equation}\centering
-\frac{\partial}{\partial\lambda}\ln{Z(\lambda)}=f_0.
\label{eq_constraint}
\end{equation}

This is a generalization of the formalism used to derive the canonical ensemble in Statistical Mechanics~\cite{Davis2016}, in which case
a uniform prior $P(\bm{x}|I_0)$ is considered, $f$ corresponds to the Hamiltonian of the system and $\lambda$ is the
inverse temperature $\beta=1/k_B T$.

In this work, we use the knowledge of maximum horizontal distance (range) $R$ of a projectile launched many times as 
$f_0 = \bar{R}$, with $f(v_0, \theta_0)=R(v_0, \theta_0)$ and by using MaxEnt we are able to determine the most unbiased
probability distribution of initial conditions $(v_0,\theta_0)$ of the projectile motion. Our results determine for the first time 
the most probably initial conditions based on the knowledge of the average range of a large number of projectile throws.
We check the validity of our analytical results by comparing with Monte Carlo sampling of compatible trajectories.

The paper is organized as follows. After this introduction, in Section \ref{methodology} we provide
a detailed analysis of the use of the maximum entropy principle and how this allows us to connect our
known information with a distributed set of probable projectile motions. Section \ref{results} presents the results of
Monte Carlo simulations used to validate our exact results. A final discussion of our results alongside insights from
the techniques presented are embodied in Section \ref{conclusions}.

\section{Methodology}
\label{methodology}

\subsection{Probability density function}

We consider the known two-dimensional motion of a projectile under the action of a constant downwards acceleration, launched
from the ground level ($y=0$) at an initial angle of $\theta_0$ with an initial speed $v_0$. The trajectory $y(x)$ of the projectile 
in the plane is given by the parabola

\begin{equation}\centering
\label{trajectory}
y(x) = x\;\tan\theta_0 -\left(\frac{g}{2v_0^2\cos^2\theta_0}\right)x^2,
\end{equation}

\noindent
with $\bm{g}=g\hat{y}$ the acceleration acting on the projectile. From here, it is straightforward that
the range, i.e. the maximum value of $x$ when touching the ground, is

\begin{equation}\centering
\label{range}
R(v_0,\theta_0)=\frac{v_{0}^2}{g}\sin{(2\theta_0)}.
\end{equation}

For a given initial angle $\theta_0$ and speed $v_0$ the kinematics of the projectile are of course deterministic.
However, a given horizontal range $R$ can be realized by an infinite combination of initial speeds and launch angles.
Accordingly, we will now consider a problem where all that is known is the individual values of $R$ for a large number 
$n$ of throws. In fact, we will only consider their average value $$\bar{R}=\frac{1}{n}\sum^{n}_{i=0} R_i$$ as known.
As $n$ becomes large we can identify $\bar{R}$ with the expected value $\langle R \rangle$ according to the law of large
numbers~\cite{Baum1965, Judd1985}. Our problem now is to find the most unbiased estimate for the probability distribution 
of initial values $\theta_0$ and $v_0$ compatible with a given $\bar{R}$, and for this we use the MaxEnt formalism.

The probability to obtain a projectile motion with initial conditions of $\theta_0$ and $v_0$ for a specific value of $\bar{R}$ 
is, in accordance with Eq. \ref{maxent}, given by

\begin{eqnarray}\centering
\label{pbb}
P(v_0, \theta_0|\bar{R}) &=&\frac{1}{Z(\lambda)}P(v,\theta|I_0)e^{-\lambda R(v,\theta)} \nonumber \\
 &=&P(v,\theta|\lambda),
\end{eqnarray}

\noindent
where $P(v,\theta|I_0)$ is the prior probability distribution, that is, the probability of these initial values without 
considering the average range. Please note that we will use $P(\ldots|\bar{R})$ and $P(\ldots|\lambda)$ indistinctly,
because $\lambda=\lambda(\bar{R})$ by Eq. \ref{eq_constraint}. The prior probability $P(v,\theta|I_0)$ is actually fixed by the
geometry of the phase space, as follows. For $\bm{v}$ in two-dimensional Euclidean space, $P(v_x, v_y|I_0)$ is a
constant, as all points in the plane are equivalent. By using polar coordinates $\bm{v}=v\hat{v}+\theta\hat{\theta}$,
we can determine the unique distribution $P(v,\theta|I_0)$ compatible with flat $P(v_x,v_y|I_0)$ as 

\begin{displaymath}\centering
 P(v,\theta |I_0)=\left<\delta(\|\bi{v}\| - v)\delta(\mathcal{A}(v_x, v_y)-\theta)\right>_{I_0},
\end{displaymath}

\noindent
where $\mathcal{A}(v_x,v_y)$ is the angle between $\bi{v}$ and the horizontal ($x$) axis. Due to the circular symmetry
of the polar representation, the angles are clearly equivalent. However, as the diameter (and therefore,
the number of points) increases linearly with $v$, the prior probability distribution of $v$ is not uniform.
Therefore, the joint prior probability distribution is described by

\begin{eqnarray}\centering
 P(v,\theta |I_0)&=&\frac{1}{k} \int_0^{v_{\textrm{max}}} dv' v' \delta(v'-v) \nonumber \\
 &=& \frac{v \Theta(v_{\textrm{max}}-v)}{\pi v_{\textrm{max}}^2},
\label{prior}
\end{eqnarray}

\noindent
by using a maximum allowed value for $v$, given by $v_\textrm{max}$. As we observe in Eq. \ref{prior}, the $\delta$ 
functions are integrated and become Heaviside step functions $\Theta$ for $v$. The final expression for the probability density is

\begin{eqnarray}\centering
P(v,\theta|I_0) & =& \frac{v\Theta(v_\textrm{max}-v)}{\Delta v_{\textrm{max}}^2} \Theta(\theta-\frac{\pi}{4}+\Delta)\nonumber \\
&&\times \Theta(\theta -\frac{\pi}{4}-\Delta),
\label{prob}
\end{eqnarray}

\noindent
where we have introduced a new parameter $\Delta$ that allows us to constrain the values of $\theta$,
avoiding the physical singularities at $\theta$=0 and $\theta=\pi/2$, where the horizontal range $R$ vanishes 
for any initial velocity $v_0$. Then, by defining $\theta$ in such a way that $\frac{\pi}{4} - \Delta<\theta<\frac{\pi}{4} + \Delta$, we 
have that the horizontal range given by Eq. \ref{range} is well-defined when $\Delta<\frac{\pi}{4}$. 
We use $\theta_{\pm}=\frac{\pi}{4}\pm\Delta$ to simplify the notation. Using this expression in (\ref{pbb}), we have

\begin{eqnarray}\centering
P(v,\theta|\lambda,I_0)&=&\frac{1}{Z(\lambda)}\Theta(v)\Theta(v_\textrm{max}
-v) \Theta(\theta_+-\theta) \nonumber \\
&&\times \Theta(\theta - \theta_-) v e^{- \frac{\lambda v^2\sin(2\theta)}{g}},
\label{model}
\end{eqnarray}

\noindent
with the partition function $Z(\lambda)$ is given by

\begin{equation}\centering
Z(\lambda)=\int_0^{{v_\textrm{max}}} dv v \int_{\theta_-}^{\theta_+} d\theta e^{- \frac{\lambda
v^2\sin(2\theta)}{g}}
\label{eq-z}
\end{equation}

Despite the possibility of solving these integrals by numerical methods, we will focus primarily in the 
analytical forms for these equations, if they exist. To obtain the distribution form, we should 
consider throws on an unlimited range for $v$, so we can take the $v_\textrm{max}\rightarrow\infty$ limit. 
By using this condition in Eq. \ref{eq-z}, we obtain an explicit form for the partition function, 

\begin{equation}\centering
\label{function-z}
Z(\lambda)=\frac{g}{4\lambda}\ln \left(\frac{\tan\theta_+}{\tan\theta_-}\right).
\end{equation}

\noindent
The value of $\lambda$ is obtained from the constraint equation (Eq. \ref{eq_constraint}) as

$$-\frac{\partial}{\partial\lambda} \ln{Z(\lambda)} = \bar{R},$$
which leads us to $\lambda=\frac{1}{\bar{R}}$. With this we have completely determined the probability density function
of initial conditions as a function of the average range, $\bar{R}$. 

By Eqs. \ref{model} and \ref{function-z}, we can determine the average trajectory, from Eq. \ref{trajectory} and the
standard assumption of $y(x)\geq 0$ for the projectile motion. Therefore, a single trajectory with the constraint on
$y(x)$ is given by

\begin{equation}\centering
y(x)=\Theta\left(\frac{v^2\sin(2\theta)}{g}-x\right)\left(\tan\theta_0 x -\frac{g}{2v_0^2\cos^2\theta_0} x^2\right).
\end{equation}

Now we can use the probability distribution in Eq. \ref{model} to compute the average trajectory of the projectile, which is
given by

\begin{equation}\centering
\label{curvesaverage}
\Big<y(x)\Big>_{\bar{R}}=\frac{\tan{\theta_+}-\tan{\theta_-}}{\ln\left(\frac{\tan\theta_+}{\tan\theta_-}\right)}\left(x e^{-x/\bar{R}}-
\frac{1}{\bar{R}}x^2\Gamma\left(0,\frac{x}{\bar{R}}\right)\right),
\end{equation}
where the notation $\Gamma\left(a,b\right)$ corresponds to the incomplete gamma function, defined by

\begin{equation}\centering
\Gamma(a,b)=\int_b^\infty dt\;e^{-t}t^{a-1}.
\end{equation}

Here we note that the average trajectory is no longer a parabola, but has long tails for large values of $x$. However it
becomes a parabola again in the limit $\bar{R} \rightarrow \infty$. Similarly we can determine the statistical average 
of the initial angle $\theta_0$, which corresponds to

\begin{equation}
\centering
\label{function_theta_av}
\left<\theta_0\right>=\frac{2}{\ln\left(\frac{\tan\theta_+}{\tan\theta_-}\right)} 
\int_{\theta_-}^{\theta_+}d\theta\frac{\theta}{\sin{ 2\theta}}.
\end{equation}

\noindent
From here it is straightforward to obtain the probability distribution of the initial angle $\theta_0$, as

\begin{equation}\centering
\label{distrib_theta}
P(\theta|\bar{R})=\frac{2}{\ln\left(\frac{\tan\theta_+}{\tan\theta_-}\right)} \frac{1}{\sin{2\theta}}.
\end{equation}

On the other hand, the statistical average of the initial velocity $v_0$ is given by

\begin{equation}\centering
\label{function_v_av}
\left<v_0\right>=\frac{4}{\bar{R}g\ln\left(\frac{\tan\theta_+}{\tan\theta_-}\right)}\int_0^\infty 
dv v^2\int_{\theta_-}^{\theta_+}d\theta e^{- \frac{v^2\sin(2\theta)}{\bar{R}g}}.
\end{equation}

Despite the fact that the integral cannot be solved analytically, we are able to identify the probability distribution 
of the velocity, which is given by

\begin{equation}\centering
\label{distrib_v}
P(v|\bar{R})=\frac{4}{\bar{R}g\ln\left(\frac{\tan\theta_+}{\tan\theta_-}\right)}v\int_{\theta_-}^{\theta_+}d\theta
 e^{- \frac{v^2\sin(2\theta)}{\bar{R}g}}.
\end{equation}

We can also compute the probability distribution of the different values of range $R$ given its average $\bar{R}$,

\begin{eqnarray}\centering
P(R|\bar{R}) & = \Big<\delta\Big(R-\frac{v^2\sin 2\theta}{g}\Big)\Big>_{\bar{R}} \nonumber \\
             & = \int_0^\infty dv \int_{\theta_-}^{\theta_+}d\theta \; \frac{1}{Z(\lambda)}v e^{-\frac{v^2\sin 2\theta}{g\bar{R}}}\delta \Big(R-\frac{v^2\sin 2\theta}{g}\Big) \nonumber \\
             & = \frac{1}{\bar{R}} e^{-R/\bar{R}},
\label{Rdist}
\end{eqnarray}

\noindent
which, interestingly, is simply an exponential distribution. This result implies that the probability of reaching or
going beyond a horizontal distance $r$ given $\bar{R}$ decreases with $r$ as

\begin{equation}
P(R \geq r|\bar{R}) = \exp(-r/\bar{R}),
\end{equation}
so it is only zero at infinity. For instance, in order to have a ``1 in 20 chance'' (5\%) of being reached by a
projectile one has to stand at a distance $r_{20} = \bar{R}\ln\;20 \sim$ 3$\bar{R}$, while $r_{1000} \sim$ 7$\bar{R}$ for a ``1 in 1000 chance'' (0.1\%).

With the distribution given by Eq. \ref{eq-z} and the statistical averages of the initial conditions of the problem (Eqs. \ref{function_theta_av} to \ref{distrib_v}), we can gain
some understanding and interpretation of the problem based on this prior information. In order to validate our analytical 
results, we numerically generated a large dataset of $(v, \theta)$ values following the probability distribution in Eq.
\ref{model} by means of the Metropolis-Hasting algorithm\cite{Gamerman2006}. With this we are be able to correlate

\begin{enumerate}
\item[(i)] angle and speed averages of the data, which we will define as the input value for an 
average conditions trajectory,
\item[(ii)] determine the probability distribution of the angles and 
velocities, to link simulations data with analytical distributions of these parameters, and 
\item[(iii)] average all trajectories generated in the distribution.
\end{enumerate}

\section{Results}
\label{results}

We present results for datasets of initial $(v,\theta)$ values that are sampled from the probability distribution
defined by Eq. \ref{eq-z} using the Metropolis-Hasting algorithm\cite{Gamerman2006}. For all the cases, we define the known 
average range $\bar{R} = 1$ m and the acceleration of gravity $g = 9.8$ m/s$^2$. The Metropolis-Hastings procedure
involved 8 million Monte Carlo steps, but only the last 320,000 where considered for production. We used a value of
$\Delta = 0.70$, close to $\pi/4 \sim$ 0.7854, allowing a broad number of $\theta$ values in the distribution. The samples were 
generated using an acceptance rate of $\sim$30\%, as usually imposed in Metropolis implementations.

\begin{figure}
\begin{center}
\includegraphics[width=0.9\columnwidth]{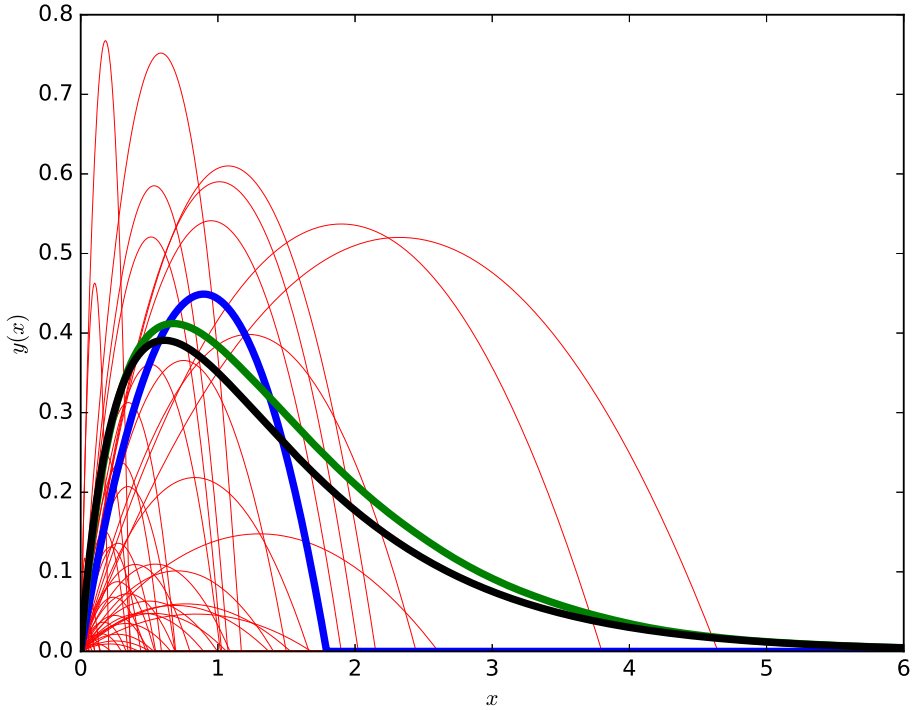}
\end{center}
\caption{Trajectories for different realizations of projectile motion. The red lines are a set of curves provided by 
equation~\ref{trajectory} with initial conditions sampled from Eq. \ref{model}. The blue line 
corresponds to Eq.~\ref{trajectory} using the average of $v_0$ and $\theta_0$ from the probability distribution. 
The black line corresponds to the average of all curves given by Eq.~\ref{curvesaverage}, meanwhile the green line 
corresponds to the average using the curves numerically sampled from the distribution.}
\label{fig:trajectories}
\end{figure}

Analysis of the probability distributions and trajectories were determined numerically, in order to compare with the analytical forms 
presented in section~\ref{methodology}. Fig. ~\ref{fig:trajectories} presents some of the generated trajectories in red. The blue 
curve corresponds to the trajectory built from the average values of $\left<\theta\right>$ and $\left< v \right>$, 
using Eq. \ref{trajectory}. We see that the obtained range using the averages of $\theta_0$ and $v_0$ is far from the
average value $\bar{R}$. Alongside this, we present the average curve according to the distribution of trajectories. As is 
expected from Eq. \ref{curvesaverage}, we observe the long-tail behavior coming from the exponential and incomplete gamma function 
$\Gamma(0,x/\bar{R})$.

\begin{figure}[h!]
\begin{center}
\includegraphics[width=0.8\columnwidth]{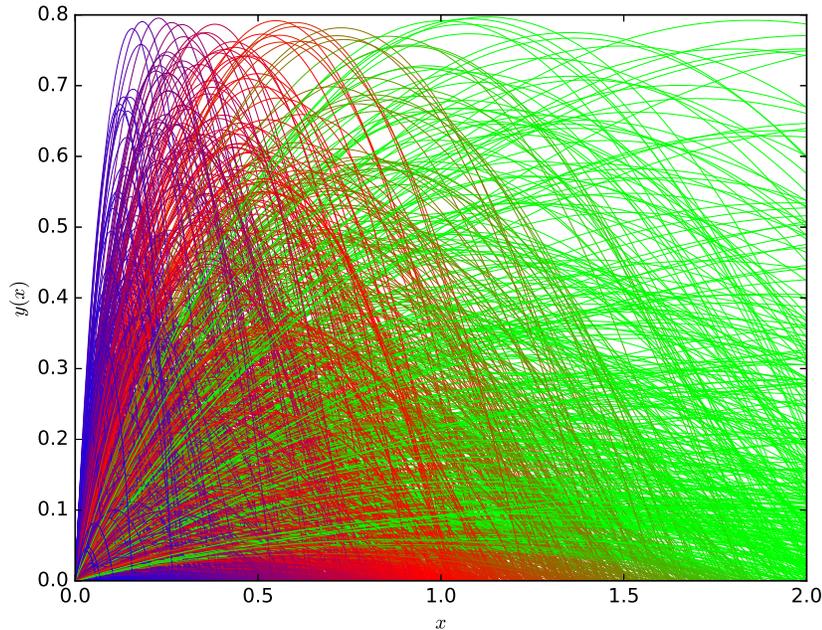}
\end{center}
\caption{Color map for a large number of trajectories, using initial conditions sampled from Eq. \ref{model}.
Values of $R$ near to 0, 1, and 2, are colored from blue, to red, to green respectively.}
\label{fig:alltraj}
\end{figure}

For the ``ideal'' case where the projectile hits right into $\bar{R}$ the maximum $y$ will always occur at $x=\bar{R}/2$. 
By using that $\left<\bar{R}\right>_{v_0,\theta_0} \sim 1.09$ we have that $\bar{R}/2 \sim 0.55$, which is close to the 
mentioned numerical value. For the case of $\bar{R}=1$, the maximum of Eqs. \ref{curvesaverage}, and~\ref{distrib_v} 
corresponds to 0.61 and 2.74 respectively, which are in excellent agreement with the data provided by the simulations
(0.65 and 2.72, respectively). A larger number of curves is displayed in Fig.~\ref{fig:alltraj}, by using a color-scale 
scheme. Here the curves are colored according to their range $R$: if $R$ is close to 0, $\bar{R}$, and 2$\bar{R}$ then the 
colors will be graduated from blue, red, and green respectively.

\begin{figure}[h!]
\begin{center}
\includegraphics[width=0.8\columnwidth]{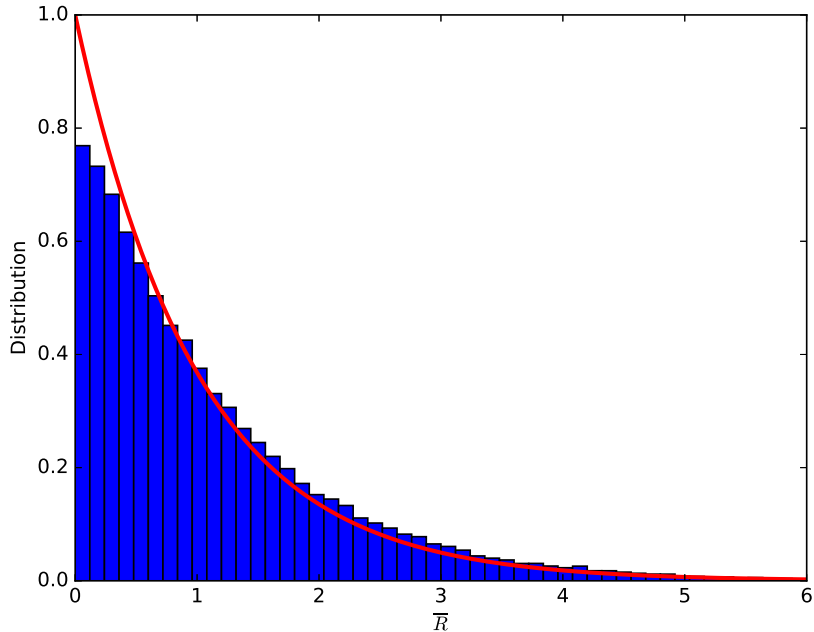}
\end{center}
\caption{$\bar{R}$ Statistical distribution of ranges $R$ from the data distribution. The solid line is the exponential distribution in Eq. \ref{Rdist}}
\label{fig:rhistogram}
\end{figure}

The histogram of $R$ compatible with $\bar{R}$ according to Eq. \ref{range} is presented in Fig. \ref{fig:rhistogram}. 
We observe a close agreement with the exponential probability distribution in Eq. \ref{Rdist} (solid line).

\begin{figure}[h!]
\begin{center}
\includegraphics[width=0.8\columnwidth]{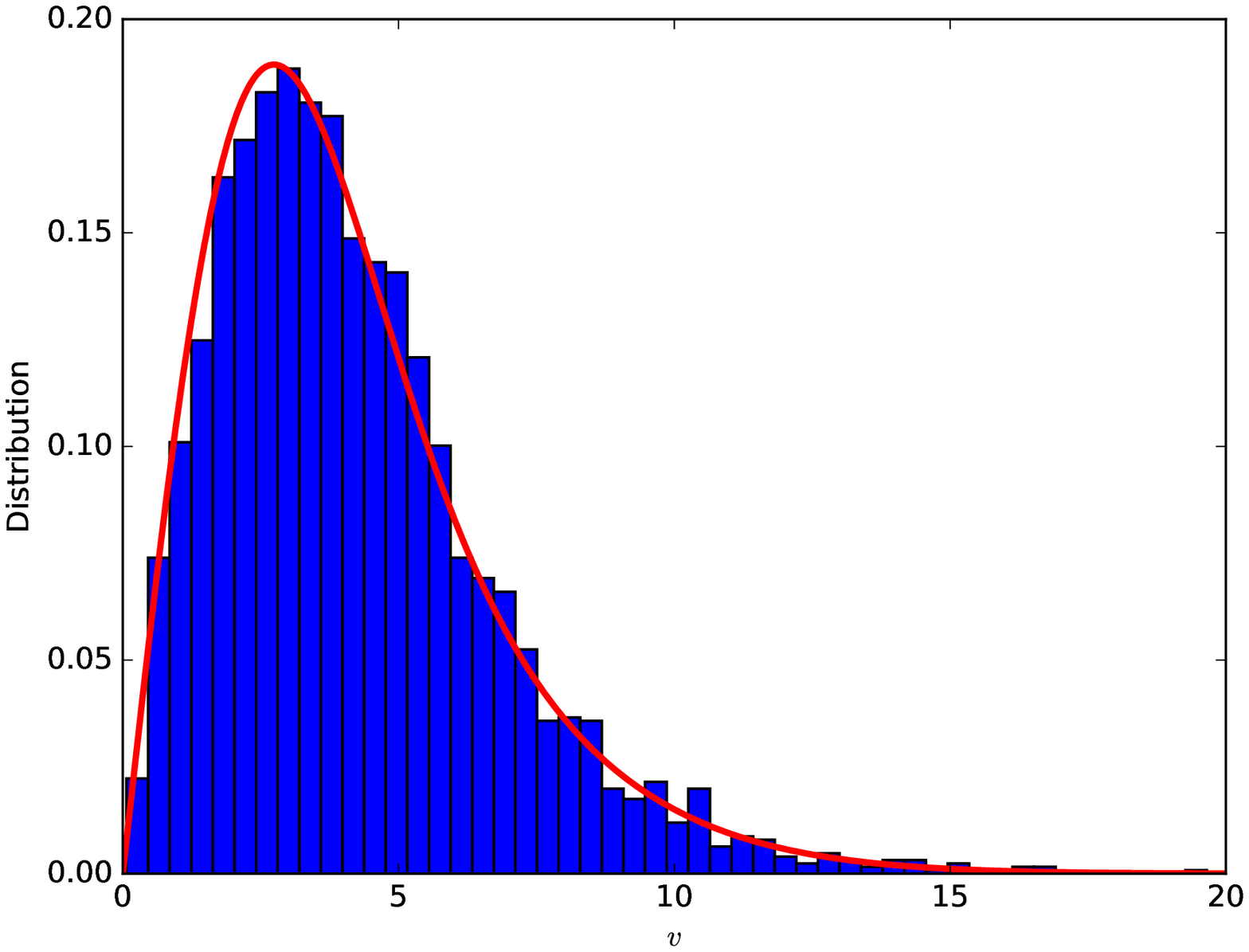}
\end{center}
\caption{$v$ Distribution of initial velocities from the generated Monte Carlo data. We observe a narrow and 
well-defined distribution centered in 2.7 m/s. The red color line, represents the analytical 
distribution computed using numerical integration, given by Eq. \ref{distrib_v}.}
\label{fig:v_histogram}
\end{figure}

Fig. \ref{fig:v_histogram} shows the histogram of the initial velocity $v_0$, having in principle values ranged from 0 to $\infty$. 
However it is important to notice that this quantity is still constrained by the average range $\bar{R}$, as is presented in Eq. \ref{distrib_v}. 
We can see a narrow and well-defined distribution centered in 2.7 m/s. The red color line represents the analytical distribution 
computed using numerical integration, given by Eq. \ref{distrib_v}. There is a good agreement between the Monte Carlo data and the predicted numerical distribution.

\begin{figure}[h!]
\begin{center}
\includegraphics[width=0.8\columnwidth]{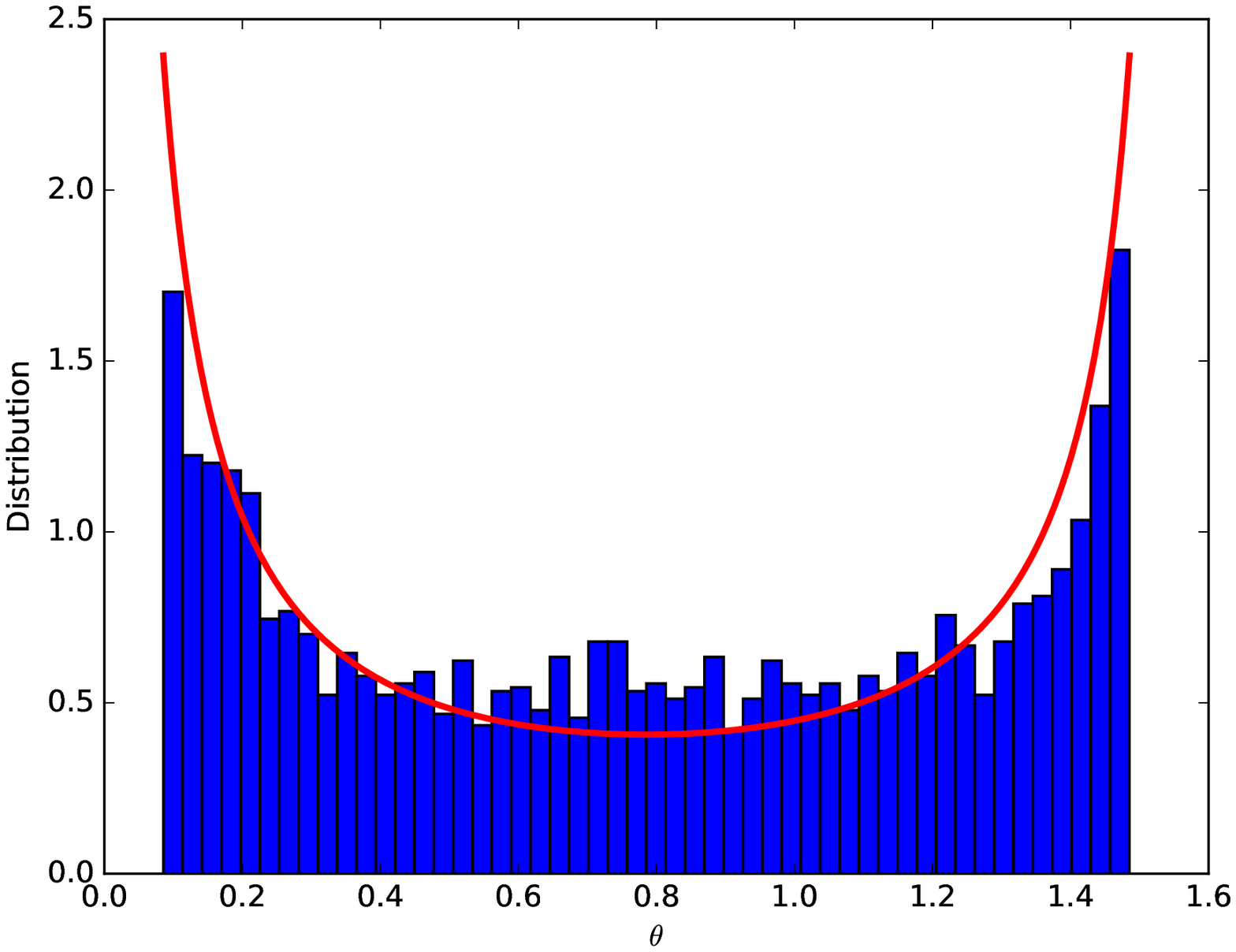}
\end{center}
\caption{$\theta$ Distribution of initial angles $\theta_0$ from the generated Monte Carlo data. The red line 
corresponds to the distribution provided by Eq. \ref{distrib_theta}.}
\label{fig:theta_histogram}
\end{figure}

As a contrast, the angle $\theta$ is more confined, for both physical and numerical reasons. From 
a physics point of view, angles larger than $\frac{\pi}{2}$ do not make sense for the projectile 
motion, and from a numerical point of view, the singularity of the solutions (Eq. \ref{prob})
is evident for the case of $\theta$=0 and $\theta = \frac{\pi}{2}$. Fig.~\ref{fig:theta_histogram} 
presents the results of the angular distribution. The red line corresponds to the distribution provided 
by Eq. \ref{distrib_theta}. A good agreement is observed in the angle distribution.

\begin{figure}[h!]
\begin{center}
\includegraphics[width=0.9\columnwidth]{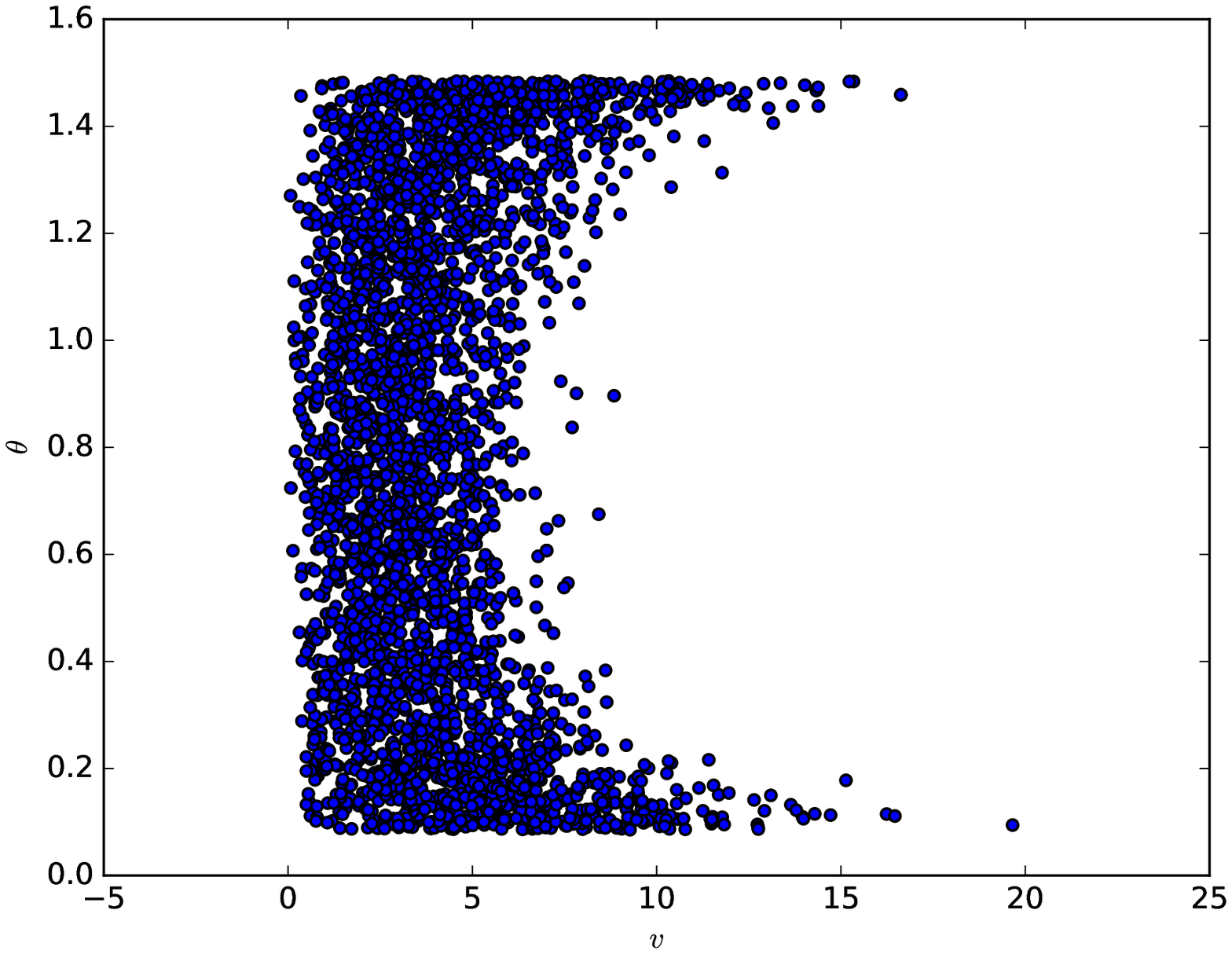}
\end{center}
\caption{$\theta$ Scatter plot of initial conditions $\theta_0$ and $v_0$. We see that for 
slow speeds the angle distribution is uniform, while for high speeds it becomes concentrated around 
$\theta$=0 and $\theta$=$\pi$/2.}
\label{fig:scattered}
\end{figure}

Another interesting point is to evaluate the possible correlations between initial speed and angle, and for this we
present in Fig. \ref{fig:scattered} the scatter plot of $(v_0, \theta_0)$ pairs. We see that for slow speeds the 
angle distribution is uniform, while for high speeds it becomes concentrated around $\theta$=0 and $\theta$=$\pi$/2. 
This is expected, as for high speeds the requirement of a low value of $\bar{R}$ constrains the angles to approach $\sin 2\theta \approx 0$.


\section{Conclusions}
\label{conclusions}
By the use of the Maximum Entropy formalism, we have shown that the problem of inferring the initial conditions for determistic physics 
problems is far from being intractable. MaxEnt allows us to provide a concise mathematical statement of the inverse
kinematic problem, which was solved both analytically and numerically via Monte Carlo simulation. We have described the
details of the procedure to infer the probability distribution of initial conditions using the projectile motion as an
example. Statistical sampling using Monte Carlo (Metropolis-Hastings) was made in order to check the validity of our
analytical results. Our results constitute a proof-of-concept of the Maximum Entropy formalism applied to solve the initial conditions 
of deterministic kinematic and dynamical problems, given information of quantities measured at posterior times.

\section*{Acknowledgements}
This work is supported by Proyecto FONDECYT 1140514. JP also acknowledges partial support from
Proyecto FONDECYT Iniciaci\'on 11130501 and Proyecto UNAB DI-15-17/RG. The authors acknowledge 
Prof. Alfonso Toro, from Universidad Andr\'es Bello, for his valuable comments to the work.

\section*{Bibliography}
\bibliographystyle{unsrt}
\bibliography{bibliography}

\end{document}